\documentclass[sigconf]{acmart}

\setcopyright{none}
\renewcommand\footnotetextcopyrightpermission[1]{}

\thanks{Copyright held by the author(s).}%

\AtBeginDocument{%
  }

\acmConference[CARS @ RecSys '25]{Workshop on Context-Aware Recommender Systems in conjunction with RecSys '25}{September 22--26, 2025}{Prague, Czech Republic}

\usepackage{bm}
\usepackage{enumitem}
\usepackage{multirow}

\DeclareMathOperator{\PR}{\mathbb{P}}
\DeclareMathOperator*{\argmax}{arg\,max}
\newcommand{\R}{\mathbb{R}}

\begin{document}

\title{Efficient and Effective Query Context-Aware Learning-to-Rank Model for Sequential Recommendation}

\author{Andrii Dzhoha}
\email{andrew.dzhoha@zalando.de}
\affiliation{%
  \institution{Zalando SE}
  \city{Berlin}
  \country{Germany}
}

\author{Alisa Mironenko}
\email{alisa.mironenko@zalando.de}
\affiliation{%
  \institution{Zalando SE}
  \city{Berlin}
  \country{Germany}
}

\author{Evgeny Labzin}
\email{evgeny.labzin@zalando.de}
\affiliation{%
  \institution{Zalando SE}
  \city{Berlin}
  \country{Germany}
}

\author{Vladimir Vlasov}
\email{vladimir.vlasov@zalando.de}
\affiliation{%
  \institution{Zalando SE}
  \city{Berlin}
  \country{Germany}
}

\author{Maarten Versteegh}
\email{maarten.versteegh@zalando.de}
\affiliation{%
  \institution{Zalando SE}
  \city{Berlin}
  \country{Germany}
}

\author{Marjan Celikik}
\email{marjan.celikik@zalando.de}
\affiliation{%
  \institution{Zalando SE}
  \city{Berlin}
  \country{Germany}
}

\renewcommand{\shortauthors}{Dzhoha et al.}

\begin{abstract}
Modern sequential recommender systems commonly use transformer-based models for next-item prediction.
While these models demonstrate a strong balance between efficiency and quality, integrating interleaving features -- such as the query context (e.g., browse category) under which next-item interactions occur -- poses challenges.
Effectively capturing query context is crucial for refining ranking relevance and enhancing user engagement, as it provides valuable signals about user intent within a session.
Unlike item features, historical query context is typically not aligned with item sequences and may be unavailable at inference due to privacy constraints or feature store limitations -- making its integration into transformers both challenging and error-prone.
This paper analyzes different strategies for incorporating query context into transformers trained with a causal language modeling procedure as a case study.
We propose a new method that effectively fuses the item sequence with query context within the attention mechanism.
Through extensive offline and online experiments on a large-scale online platform and open datasets, we present evidence that our proposed method is an effective approach for integrating query context to improve model ranking quality in terms of relevance and diversity.
\end{abstract}

\ccsdesc[500]{Information systems~Recommender systems}

\keywords{Sequential Recommender Systems, Retrieval, Query Context, Transformers}

\maketitle

\section{Introduction}

Modeling sequential interactions is essential in applications like e-commerce, social media, and streaming services, where historical user behaviors inform future recommendations.
A well-designed sequential recommender system can drive significant product and business impact by surfacing relevant items at the right time.
To achieve this, modern recommender systems typically frame their objective as next-item prediction, with transformer-based models emerging as the state-of-the-art approach \cite{Kang2018SelfAttentiveSR, 10.1145/3357384.3357895, 10.1145/3336191.3371786}.

\begin{figure}[t]
  \centering
  \includegraphics[width=1.0\linewidth]{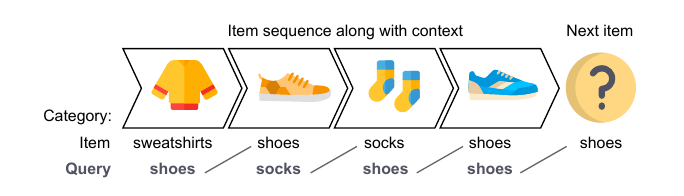}
  \caption{
    A user's journey is defined by the items they interacted with and the context of those interactions.
    The query category refers to the category under which the user engages with an item, while the item's category is part of its features.
    Each next item follows a previous item interaction and a query context change.
    The model predicts recommendations based on past actions and a typically feature-rich query context, including search queries, product categories, time of day, weather, season, and location.
  }
  \label{fig:item-sequence}
  \Description[The user's intent is a combination of past actions and the current query]{
    The recommender system predicts the next item to recommend based on the user's journey,
    along with the current query request, such as the category they are currently browsing.
  }
\end{figure}

User behavior in recommender systems is substantially influenced by intent, and \textit{query context}~\cite{10.1145/1401890.1401995}
-- such as browse category, search query, landing page, etc., -- serves as a strong indicator of that intent within a session.
For example, if a user is browsing the "shoes" category in an e-commerce store or "comedy movies" on a streaming platform,
their next action is likely to involve selecting an item from that category.
Query context can include factors like search queries, product categories, time of day, weather or season, and location, all of which provide valuable signals for recommendation.
This context is particularly useful in cold-start scenarios, where there is limited historical interaction data.
Figure~\ref{fig:item-sequence} illustrates the query context for a given item sequence.

However, integrating query context into sequential recommendation models presents several challenges.
First, privacy restrictions and feature store limitations \cite{pmlr-v67-li17a}, can prevent online access to historical query context, which can be highly feature-rich.
As a result, models must rely on a user's past interactions and limited query context (e.g., their current query context) available during inference,
which can lead to a discrepancy between training and serving environments.
Second, technical difficulties arise when incorporating query context into transformer-based models.
Popular sequential models like SASRec~\cite{Kang2018SelfAttentiveSR} and BERT4Rec~\cite{10.1145/3357384.3357895} use attention mechanisms to efficiently predict multiple next items in a single forward pass during training,
either through causal language modeling or masked language modeling.
Unlike an item's features, which can also serve as contextual information, query context is not temporally aligned with the item sequence, as it is shifted and corresponds to the next item.
This makes integration into the model's attention flow more challenging, as it requires additional design considerations to prevent training-serving mismatches and ensure proper propagation of the query context across layers.

Existing research on efficient transformer-based sequential models primarily focuses on integrating an item's features as contextual information \cite{YANG2021157, 10.1145/3404835.3462832}
or treating query context as side information in conjunction with the user representation,
without fusing it early with item sequence \cite{10.1145/2959100.2959190, 10.1145/3219819.3219823},
which may limit model effectiveness.
When historical query context is assumed to be available online, it has been effectively incorporated
into recurrent neural network-based approaches by embedding it within the item sequence,
thereby fusing it with the user representation (\cite{7837948, 7917325, 10.1145/3159652.3159727}).

In this paper, we propose an efficient method to integrate query context into transformer-based sequential models, even when historical context is unavailable at inference.
Specifically, we address two research questions:
\begin{itemize}
  \item RQ1: How can query context be effectively integrated into transformer-based sequential recommendation models while mitigating the training-serving mismatch?
  \item RQ2: Does the absence of historical query context during inference negatively impact model performance?
\end{itemize}
To answer these, we explore various approaches for incorporating query context into transformer-based sequential recommenders,
using causal language modeling as a case study. Our key contributions include a comprehensive study on efficient and effective query context integration,
a method that enhances user representation by contextually fusing query context within transformer-based ranking models to improve accuracy,
and extensive offline and online experiments on a large-scale platform, validated with open datasets and supported by open-source code for reproducibility.

\section{Related work}

The introduction of the transformer architecture in \textit{Attention Is All You Need}~\cite{10.5555/3295222.3295349} revolutionized deep learning by significantly improving training efficiency and addressing long-range dependency challenges.
Transformer-based models, leveraging self-attention mechanisms, have become the state-of-the-art in learning-to-rank tasks across both retrieval and ranking stages in recommender systems \cite{celikik2024buildingscalableeffectivesteerable, Kang2018SelfAttentiveSR, 10.1145/3357384.3357895, dzhoha2024reducingpopularityinfluenceaddressing}.

Recent work in sequential recommendation has primarily focused on short-term predictions, in particular the next-item prediction task \cite{10.1145/3190616}.
Popular models such as SASRec~\cite{Kang2018SelfAttentiveSR}  and BERT4Rec~\cite{10.1145/3357384.3357895} use attention mechanisms to predict multiple future interactions in a single forward pass during training, using either causal language modeling or masked language modeling.
However, integrating interleaving features, such as query context, poses challenges since they are not temporally aligned with the item sequence fed into the attention mechanism.

Much of the literature on context-aware recommendations focuses on enriching an item's representation with its features as contextual information \cite{8031316, YANG2021157, 10.1145/3404835.3462832},
often under the explicit assumption that the query context -- i.e., the conditions under which a user takes the next action -- is not observable by the system \cite{10.1145/3404835.3462832, Adomavicius2015}.
Contextual features such as brand, title, category, and price play a crucial role in recommendation quality.
These features help generalization and improve recommendations for long-tail items (those with fewer observations).
Enhancing item representations with these features \cite{YANG2021157} or separately modeling item-to-item and context-to-context transitions within attention-based models \cite{10.1145/3404835.3462832}
has been shown to improve accuracy, particularly in sparse data settings.
Since the contextual features are temporarily aligned with the item sequence, incorporating them into the input of transformer-based models does not pose a challenge.

Query context is typically incorporated as side information alongside the user representation,
rather than being fused early with the item sequence \cite{10.1145/2959100.2959190, 10.1145/3219819.3219823}.
However, treating query context as side information may limit the predictive power of the user representation,
as it is not explicitly tailored to the query context.

Alternatively, query context can be integrated into an item's representation by aligning it with the item sequence in the attention mechanism.
While this early fusion improves performance (\cite{10711191, 10.1145/3477495.3531963}), it has drawbacks.
Placing query context in the attention query guides fusion but also misplaces next-item context in key/value positions, complicating learning.
Moreover, it assumes historical query context is always available, which is often infeasible due to privacy and scalability constraints (\cite{pmlr-v67-li17a}).
In this paper, we address both of these issues in detail through our analysis and experiments.

With this study, we aim to bridge this gap by effectively integrating query-contextual information into attention-based sequential models
trained with a causal language modeling procedure as a case study.

\section{Methodology}

\subsection{Problem statement}

We address the sequential recommendation problem, where a model predicts the next item a user is likely to interact with based on their historical interactions.
Formally, let $u \in \mathcal{U}$ be a user with an interaction history represented as a sequence of items $\left(x_i^{u}\right)_{i=1}^{N+1}$ arranged in chronological order.
Given this sequence and the contextual query information $c_{i+1}^{u}$ at the next time step, the goal is to estimate the probability distribution over all candidate items $x \in \mathcal{X}$:
\begin{equation*}
  \PR\left(x_{i+1}^{u} = x \;\middle|\; x_1^{u},\dots,x_i^{u};\,c_{i+1}^{u}\right).
\end{equation*}
Here, $N$ is the number of items the user interacted with.
During estimation, the $(N+1)$th item is used only as the target when predicting the $N$th next item,
while the first item prediction is used for cold-start users.

\subsection{Method}\label{sec:method}

To efficiently model the problem, we consider the deep self-attention transformer model trained with causal language modeling,
following the SASRec-style architecture \cite{Kang2018SelfAttentiveSR, 10.1145/3336191.3371786, 10.5555/3367471.3367642}.
A transformer begins with an item embedding, followed by a series of self-attention layers.
These models stack multiple self-attention layers, each followed by a position-wise feed-forward sublayer, residual connections, and layer normalization \cite{ba2016layer, nguyen2019transformers}.
Multi-head self-attention is commonly used to capture different aspects of input sequences \cite{10.5555/3295222.3295349},
though we omit it here for notational simplicity.

\paragraph{Self-attention layer blocks}
Given an input $\bm{X}^{(h)} \in \R^{N \times D}$ to the $h$th layer block,
the hidden states in the output layer $\bm{X}^{(h+1)}$ are computed by attending to the states of $\bm{X}^{(h)}$.
The $i$th row vector of $\bm{X}^{(h)}$ is denoted as the $1 \times D$ matrix $\bm{X}_{i}^{(h)}$.
The input $\bm{X}_{i}^{(1)}$ represents the embedded representation of item $x_i$ with a dimension of $D$.

\begin{figure}[t]
  \centering
  \includegraphics[width=0.9\linewidth]{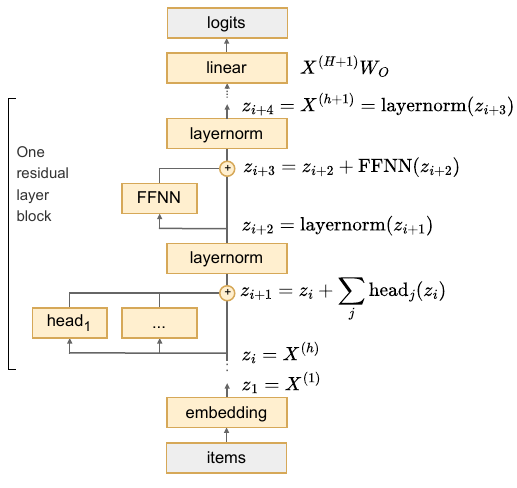}
  \caption{
    High-level architecture of the transformer with $H$ residual layers, each including layer-specific multi-head attention ($\mathrm{head}_j$) and a feedforward neural network ($\mathrm{FFNN}$).
  }
  \label{fig:transformer}
  \Description[High-level architecture of the transformer]{
    High-level architecture of the transformer, consisting of several stacked residual layer blocks, including attention heads and a feedforward neural network.
  }
\end{figure}

Specifically, the input layer $\bm{X}^{(h)}$ is first transformed into the query, key, and value matrices
$\bm{Q}^{(h)}, \bm{K}^{(h)}, \bm{V}^{(h)} \in \R^{N \times D}$:
\begin{equation}\label{eq:qkv-projections}
  \begin{bmatrix}
    \bm{Q}^{(h)}\\
    \bm{K}^{(h)}\\
    \bm{V}^{(h)}
  \end{bmatrix}
  =
  \bm{X}^{(h)}
  \begin{bmatrix}
    \bm{W}_{Q}^{(h)}\\
    \bm{W}_{K}^{(h)}\\
    \bm{W}_{V}^{(h)}
  \end{bmatrix},
\end{equation}
where $\bm{W}_Q^{(h)}, \bm{W}_K^{(h)}, \bm{W}_V^{(h)} \in \R^{D \times D}$ are layer specific parameter matrices.
The attention mechanism computes compatibility scores via a dot-product operation between the query and key matrices:
\begin{equation}\label{eq:compatiblity-score}
  \bm{Q}^{(h)} \left(\bm{K}^{(h)}\right)^{\intercal}.
\end{equation}
The scores are then normalized by $\sqrt{D}$, followed by a row-wise softmax operation, which is applied to the value matrix.
The output layer $\bm{X}^{(h+1)}$ is then computed as
\begin{equation*}
  \bm{X}^{(h+1)} = g^{(h)}\left(
  \bm{X}^{(h)},
  \,\mathrm{softmax}
  \left( \bm{Q}^{(h)} \left( \bm{K}^{(h)} \right)^{\intercal} / \sqrt{D} \right) \bm{V}^{(h)}
  \right),
\end{equation*}
where $g^{(h)}\left(z, z'\right)$ encapsulates additional transformations of the attention output $z'$,
such as feed-forward sublayer projection, dropout, and layer normalization,
while also incorporating residual connections with $z$.
All these additional transformations are applied position-wise along the last dimension (the embedding dimension).
The complete architecture is illustrated in Figure~\ref{fig:transformer}.

\paragraph{Causal language modeling}
As in \cite{Kang2018SelfAttentiveSR}, to efficiently predict each next item $x_{i+1}$ within a user sequence
while reusing all the previous states $\left(X_t^{(h)}\right)_{t \leq i}$ in a single forward pass,
we modify the attention mechanism in Eq.~(\ref{eq:compatiblity-score}) by restricting connections between $\bm{Q}_i$ and $\bm{K}_j$ for all $j > i$.

\paragraph{Training procedure}
Given a user sequence $\left(x_i^u\right)_{i=1}^{N}$ represented by $\bm{X}^{(1)}$, the model
processes it through multiple transformations layers $\left(g^{(h)}\right)_{h=1}^{H}$,
predicting the relevance scores $y_{i, j}^{u}$ for each candidate item $x_j \in \mathcal{X}$
given the past interactions $\left(x_t^u\right)_{t=1}^{i}$:
\begin{equation}\label{eq:rel-score}
  y_{i, j}^{u} = \bm{X}_{i}^{(H+1)} \bm{W}_O \bm{A}_{j}^{\intercal},
\end{equation}
where $\bm{W}_O \in \R^{D \times D}$ is the final output projection matrix and $\bm{A} \in \R^{|\mathcal{X}| \times D}$ is the learnable item embedding matrix.

The model is trained using categorical cross entropy loss \cite{di2024theoretical}:
\begin{equation*}
  \sum_{u \in \mathcal{U}} \sum_{i=1}^{N} \log\left(
  \sum_{j \in \mathcal{X} /\ \left\{x_{i+1}^{u}\right\}} \exp\left(y_{i, j}^{u} - y_{t, x_{i+1}^u}^{u} \right) + 1
  \right),
\end{equation*}
where $x_{i+1}^u$ represents the target item given $\left(x_t^u\right)_{t=1}^{i}$.

To handle variable-length sequences, padding tokens are added to the left and masked during both attention computation (\ref{eq:compatiblity-score}) and loss computation.

\paragraph{Inference}
During inference, the position of the last item is used to predict the most probable next item given $\left(x_i^u\right)_{i=1}^{N}$:
\begin{equation*}
  \argmax_{j=1,\dots,|\mathcal{X}|} \left\{\, y_{N, j}^{u} \in \R : y_{N, j}^{u} = \bm{X}_{N}^{(H+1)} \bm{W}_O \bm{A}_{j}^{\intercal} \,\right\},
\end{equation*}
or to rank a subset of items based on their dot-product relevance scores $\left( y_{N, j}^{u} \right)_j$.

\subsection{Query context integration}\label{sec:query-context-integration}

Due to causal masking, self-attention layers enable fully parallelized next-item predictions within a user sequence in a single forward pass.
The computational complexity is $O\left( N^2 D + N D^2 \right)$, where the first term arises from self-attention and the second from the feed-forward network.
This is significantly more efficient than recurrent neural network-based methods while maintaining high predictive accuracy \cite{Kang2018SelfAttentiveSR, 10.1145/3336191.3371786, 10.5555/3367471.3367642}.
Our goal is to improve recommendation accuracy by incorporating query context while maintaining computational efficiency.
We hypothesize that query context can:
\begin{enumerate}
  \item Guide user representation by incorporating global constraints (e.g., prioritizing shoe-related items when searching for footwear).
  \item Refine user representation, making it more tailored to the query context.
  \item Improve cold-start recommendations by leveraging query context to identify the most relevant and popular items in cases of sparse user interactions.
\end{enumerate}

\section{Query context temporal misalignment}

Such sequential recommender models optimize efficiency by computing losses for all next-item predictions in a single forward pass.
However, this introduces a challenge: query context features become temporally misaligned.
To predict the next item $x^u_{i+1}$, the loss is computed at the $i$th hidden output position $\bm{X}_{i}^{(H+1)}$.
Thus, the query context $c^u_{i+1}$ must be integrated at the attention's query position $i$ to properly fuse historical information.

\subsection{Fixing temporal misalignment}\label{sec:fixing-temporal-misalignment}

A straightforward alignment method shifts the query context matrix $\bm{C} \in \R^{(N +1) \times D}$
one step back so that aligns $c^u_{i+1}$ with $\bm{X}_{i}^{(H+1)}$.
By doing so, we can integrate the query context as contextual information in the input by concatenating it with the corresponding item embeddings, similar to \cite{YANG2021157, 10.1145/3404835.3462832}.

\begin{figure}[t]
  \centering
  \includegraphics[width=1.0\linewidth]{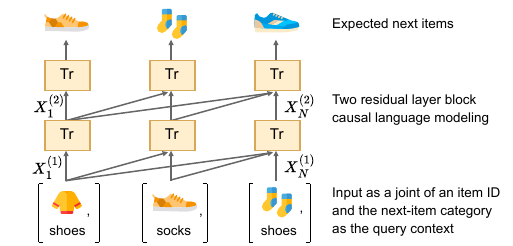}
  \caption{
    An illustration of the attention flow among positions ($\mathrm{Tr}$) and layers when aligning query context information and placing it in the input.
    When only the current query context is available online ("shoes" at the last position), past actions will have padding in the input context embeddings, creating a training-serving mismatch.
  }
  \label{fig:clm}
  \Description[High-level architecture of the transformer]{
    An illustration of the attention flow among positions (Tr) and layers when aligning query context information and placing it in the input.
  }
\end{figure}

Specifically, we can modify the transformer's input by adding shifted query context embeddings to item embeddings in the first layer's query, key, and value projections in Eq.~(\ref{eq:qkv-projections}):
\begin{equation}\label{eq:first-layer-proj}
  \begin{bmatrix}
    \bm{Q}^{(1)}\\
    \bm{K}^{(1)}\\
    \bm{V}^{(1)}
  \end{bmatrix}
  =
  \left(\bm{X}^{(1)} + \bm{L}\bm{C} \right)
  \begin{bmatrix}
    \bm{W}_{Q}^{(1)}\\
    \bm{W}_{K}^{(1)}\\
    \bm{W}_{V}^{(1)}
  \end{bmatrix},
\end{equation}
where $\bm{L}$ is a modified lower shift binary $N$-by-$(N+1)$ matrix with ones on the subdiagonal.
The projections (\ref{eq:qkv-projections}) in subsequent layers $h = 2,\dots,H$ remain unchanged.
This ensures $c_{i+1}$ is correctly placed at query position $\bm{Q}^{(1)}_{i}$,
influencing attention over past items $\left(x_t^u\right)_{t=1}^{i}$
while also propagating the query context further through keys and values.
However, this approach raises concerns:
\begin{enumerate}
  \item \textbf{Training-serving mismatch:}
  When historical query context is absent online, it remains in keys and values during training, creating a mismatch between training and inference.
  \item \textbf{Complex learning dynamics:}
  Instead of directly attending to past items, attention focuses on representations conditioned on next-item query contexts.
  This may hinder learning both within a single attention layer and across layers, as hidden outputs are reused causally for efficiency.
  \item \textbf{Misleading representations:}
  Each past item is represented alongside the next-item query context rather than its own intrinsic features.
  For example, a sneaker might be paired with the "socks" category if the next action involved socks, rather than retaining its true "shoes" category.
  This can unnecessarily complicate learning or reduce generalization, particularly in sparse data settings.
  \item \textbf{Limited extensibility:}
  Methods like random time window masking \cite{10.1145/3580305.3599918}, which randomly mask past items to improve generalization and long-term user retention,
  also remove query context when it is embedded in item representations, leading to unnecessary information loss.
\end{enumerate}
The attention flow of this approach is shown in Figure~\ref{fig:clm}.

\subsection{Less efficient query context integration}

Before exploring efficient approaches, we first consider a simpler but computationally expensive method that fully satisfies the assumptions in
Section~\ref{sec:query-context-integration} without introducing the concerns previously mentioned.
Instead of computing all next-item predictions in a single forward pass, we can split each user sequence incrementally by constructing a set of subsequences:
\begin{equation*}
  \left\{\, \left( x_{t}^{u} \right)_{t=1}^{i} : i = 1,\dots,N \,\right\}.
\end{equation*}
Each subsequence is padded on the left as described earlier, and we perform a forward pass separately for each,
focusing on predicting only the next item at the final position while masking the rest in the loss computation.

Since we are only interested in the final output of the transformer, $\bm{X}_{N}^{(H+1)}$,
incorporating the query context is straightforward: we fuse the embedding of the last item in each subsequence, $x_i$,
with the corresponding query context $c_{i+1}$ embedding at the final input position $\bm{X}_{N}^{(1)}$.
This ensures that the query context is placed at the attention query position, guiding attention over past interactions
while also propagating the query context information through the key and value representations further.

However, this method significantly increases computational complexity to $O\left( N^3 D + N^2 D^2 \right)$.
This cubic complexity makes it less practical for large-scale applications, motivating the need for more efficient approaches,
which we explore in the following sections.

\section{Approach}\label{sec:approach}

Before integrating the query context, we first establish the technical specifications it must fulfill.
Refining our assumptions from Section~\ref{sec:query-context-integration},
we define the expressiveness of the query context in three distinct roles:
\begin{itemize}[leftmargin=1em]
  \item [] \textbf{As a separate feature:}
  \begin{itemize}[leftmargin=2em]
    \item [(1)] To help guide the customer representation in conjunction with the query context.
    \item [(2)] To identify popular items within the query context, particularly for cold-start users.
  \end{itemize}
  \item [] \textbf{As part of the attention mechanism:}
  \begin{itemize}[leftmargin=2em]
    \item [(3)] To refine the customer representation, making it more tailored to the query context.
  \end{itemize}
\end{itemize}

\subsection{Approach A: Query context outside}
To address the first two requirements, we integrate the query context $\bm{C}$ as a separate feature.
We modify the method in Section~\ref{sec:method} by adding a shifted context matrix with the output of the transformer
when predicting relevance scores $y_{i, j}^{u}$ for each candidate item $x_j \in \mathcal{X}$
given past interactions $\left(x_t^u\right)_{t=1}^{i}$ in Eq.~(\ref{eq:rel-score}):
\begin{equation}\label{eq:context-outside-addition}
  y_{i, j}^{u} = \left(\bm{X}_{i}^{(H+1)} \bm{W}_O + \left(\bm{L}\bm{C}\right)_i \right) \bm{A}_{j}^{\intercal}.
\end{equation}
This approach avoids the training-serving mismatch when historical query context is unavailable online
and does not introduce any of the concerns mentioned in Section~\ref{sec:fixing-temporal-misalignment}.
Its limitation is that it doesn't fuse the past items with respect to the query context within the attention mechanism.

\subsection{Approach B: Query context in the input}\label{sec:approach-b}
To incorporate all three aspects of query context integration, we apply the approach from Section~\ref{sec:fixing-temporal-misalignment} by modifying the transformer's input.
We add shifted query context embeddings to item embeddings in the first layer (Eq~(\ref{eq:first-layer-proj})).
To partially address the concerns outlined in Section~\ref{sec:fixing-temporal-misalignment},
we randomly mask the query context information with padding with a probability optimized via grid search (i.e., query context masking).

Despite these concerns, we evaluate this approach because it offers a straightforward way to incorporate query context
while maintaining compatibility with similar standard transformer implementations.
It requires minimal modifications, making it a practical starting point for fusing past items with query context,
particularly when using off-the-shelf transformer implementations.

\subsection{Approach C: Query context in the last layer's query position and outside}
\begin{figure}[t]
  \centering
  \includegraphics[width=1.0\linewidth]{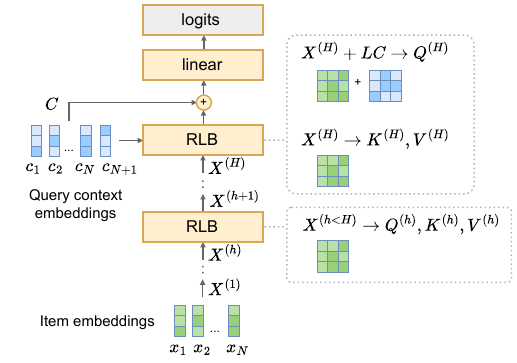}
  \caption{
    Approach C integrates query context into the transformer with residual layer blocks ($\mathrm{RLB}$),
    adding it at the query position in the last layer and summing it with the output.
  }
  \label{fig:approach-c}
  \Description[High-level architecture of the transformer]{
    Query information is added at the query position in the last layer and concatenated with the output of the transformer.
  }
\end{figure}
Analyzing attention flow in causal language modeling, we conclude that the safest way to introduce query context within attention
without causing unintended misrepresentations (Section~\ref{sec:fixing-temporal-misalignment})
is by placing it in the last layer, exclusively at the query position \cite{2401.09343}.

If the query context were introduced in key and value positions at any layer,
it would cause next-item information to propagate through subsequent attention layers, violating the training-serving consistency or causing unintended misrepresentations.
Similarly, introducing it in the query positions of earlier layers would condition hidden states $\bm{X}^{(h)}_i$ on next-item representations.
These conditioned states would then be reused in later layers and positions, $\bm{X}^{(h'>h)}_{i'\geq i}$, due to causal language modeling.

To address these concerns, we modify the method from Section~\ref{sec:method}
by adding shifted query context embeddings to the last layer's query position without modifying key or value projections:
\begin{equation*}
  \begin{aligned}
    \bm{Q}^{(H)} &= \left(\bm{X}^{(H)} + \bm{L}\bm{C} \right) \bm{W}_{Q}^{(H)}, \\
    \begin{bmatrix}
      \bm{K}^{(H)}\\
      \bm{V}^{(H)}
    \end{bmatrix}
    &=
    \bm{X}^{(H)}
    \begin{bmatrix}
      \bm{W}_{K}^{(H)}\\
      \bm{W}_{V}^{(H)}
    \end{bmatrix},
  \end{aligned}
\end{equation*}
while the projections in the previous layers, $1 = 2,\dots,H-1$, remain unchanged as defined in Eq.~(\ref{eq:qkv-projections}).

In this way, past item representations are fused with query context while preserving proper attention flow.
However, this method alone does not fully satisfy the first two requirements (separate query context as a feature),
which limits its effectiveness for cold-start customers.
To compensate for this limitation, we also add a shifted context matrix to the transformer's output,
as done in Approach~A (Eq.~(\ref{eq:context-outside-addition})).

This final implementation ensures that:
\begin{itemize}
  \item The past item representations are fused with query context in attention.
  \item Query context is explicitly provided as a feature, benefiting cold-start scenarios.
  \item The approach remains robust to training-serving mismatches and other concerns outlined in Section~\ref{sec:fixing-temporal-misalignment}.
\end{itemize}
A high-level diagram of this approach is illustrated in Figure~\ref{fig:approach-c}.

\section{Experiments on our large-scale online platform}

In this section, we present the experimental results of incorporating query context information using the three approaches introduced earlier.
We integrate query context information into the ranking model that powers both the Browse and Search use cases of our e-commerce platform's catalog.

\paragraph{Base model}
The model follows a two-tower architecture \cite{10.1145/2959100.2959190, twotowersampling}, with the training procedure described in Section~\ref{sec:method}.
It consists of a user tower, represented by a transformer with residual layer blocks, and an item tower,
represented by a feed-forward network, whose output is the item embedding $\bm{A}$ from Eq.~\ref{eq:rel-score}.
Although these towers are trained jointly, they are deployed and operated independently.
The item embeddings are indexed in a vector store for efficient ranking and retrieval.
Each time a user makes a request, the user tower generates a new user embedding based on their most recent interactions (up to the last 100 items).
This embedding is then used to retrieve the most relevant items using a nearest-neighbor index.

During training, we follow the approach described in Section~\ref{sec:method}, employing a sampled softmax loss with log-uniform sampling.
Negative classes make up 0.5\% of all classes, a ratio fine-tuned offline to balance training efficiency and effectiveness.
We use two residual layer blocks and a four-head multi-head attention setting.
We represent an item input by concatenating the embeddings of interacted item, interaction type, and categorical timestamp data.

\paragraph{Dataset}
Our dataset consists of item interactions from the past 60 days, aggregated by user ID.
These interactions include item clicks, add-to-wishlist, add-to-cart, and checkouts events,
all attributed to the Browse and Search use cases using a "last-touch" attribution model.
Each interaction is joined with the corresponding item ID, timestamp, and interaction type, then sorted chronologically.
The training dataset includes over 70 million unique users across 25 markets, while the evaluation dataset contains more than 500,000 users.
To prevent data leakage, we apply a strict temporal split between the training and test datasets.

\paragraph{Methodology}
To enhance ranking performance, we incorporate query context information as a proxy for customer intent.
This includes a browsing category, search query (if present), and other relevant user attributes available during inference.
However, only the current query context is accessible at inference time, whereas historical query context is available during training.
For offline evaluation, we primarily rely on the following metrics~\cite{10.1145/3460231.3478848}:
\begin{itemize}
  \item Recall@k: The proportion of all relevant items within the top-k retrieved items.
  \item NDCG@k: A ranking effectiveness measure that considers both item relevance and position within the top-k ranked results, where attributed items are treated as relevant.
\end{itemize}

\subsection{Offline experiments}

We offline evaluate the model on a holdout set containing instances from the subsequent day,
ensuring that the evaluation data remains unseen during the training phase.
We compute metrics for each next item prediction within the holdout set day.
This resembles well our production setup as we train a model on a daily basis.

Table~\ref{tab:zalando-offine} presents the offline evaluation results of the three approaches described in Section~\ref{sec:approach}.
As shown, incorporating query context as a separate feature (Approach A) improves performance in both use cases.
Furthermore, fusing past items based on the query context (Approaches B and C) leads to even greater performance gains,
indicating that refining user representation to better align with the query context is beneficial.
While the Browse use case shows greater improvements, we note that our search query currently has a limited categorical representation of the query context,
and we plan to enhance it in the future.

It is worth noting that Approach C shows the highest improvement, particularly in NDCG.
We hypothesize that this improvement stems from properly handling the training-serving mismatch and avoiding misleading representations.

\begin{table}[t]
  \caption{
    Offline evaluation results of integrating query context into the ranking model that powers both the Browse and Search use cases of our e-commerce platform's catalog.\\
    The table presents the relative uplift of three approaches:\\(A) query context outside, (B) query context in the input with masking 0.3, (C) query context in the last layer and outside.
  }
  \label{tab:zalando-offine}
  \centering
  \begin{tabular}{l l c c c}
    \toprule
    & & \multicolumn{3}{c}{\textbf{Approach}} \\
    \cmidrule(lr){3-5}
    \textbf{Use case} & \textbf{Metric} & A & B & C \\
    \midrule
    \multirow{2}{*}{Browse}
    & Recall@500 & +2.81\% & +4.69\% & +4.66\% \\
    & NDCG@500   & +3.03\% & +5.61\% & +6.56\% \\
    \midrule
    \multirow{2}{*}{Search}
    & Recall@500 & +1.48\% & +2.27\% & +2.32\% \\
    & NDCG@500   & +1.44\% & +2.23\% & +3.04\% \\
    \midrule
    \multirow{2}{*}{All}
    & Recall@500 & +2.11\% & +3.41\% & +3.41\% \\
    & NDCG@500   & +2.17\% & +3.78\% & +4.65\% \\
    \bottomrule
  \end{tabular}
\end{table}

\subsection{Online experiments}

In our initial iterations of integrating query context, we evaluated Approaches A and B using an off-the-shelf transformer implementation (Section~\ref{sec:method}).
This led to an online A/B test on the Browse and Search ranking use cases, focusing on Approach B -- query context in the input.
Within this approach, we fine-tuned the query context masking (Section~\ref{sec:approach-b}) offline by randomly omitting query context information per position.
We optimized the masking rate to 30\% via grid search, selecting the highest value that did not cause significant offline performance deterioration.

We allocated equal traffic splits among variants over several weeks to achieve the minimum detectable effect for the success KPI, with a p-value < 0.05.
The main A/B test results for incorporating query context information using Approach B are presented in Table~\ref{tab:zalando-online}.
The results demonstrate significant positive uplifts in both engagement and financial metrics.
Below is a summary of key insights from a deep dive into the exploratory KPIs:
\begin{enumerate}
  \item For high-value actions (add-to-wishlist and add-to-cart), both new and returning users saw positive impacts:
  +1.0\% [0.2, 1.8]\% for new and +1.7\% [1.4 , 2.0]\% for returning users.
  This suggests that the treatment benefits even users for whom we have little to no prior data.
  \item Order-related metrics showed clear improvements: the number of items per user increased by +0.7\% [0.3, 1.0]\%,
  order per user by +0.6\% [0.3, 0.8]\%, and the percentage of users with at least one order by +0.6\% [0.3, 0.8]\%.
  This suggests the treatment boosted purchasing activity.
  \item The average discovery return days per user improved by +0.2\% [0.1, 0.3 ]\%.
  \item The ranking output became more diverse.
  For example, impressions on non-diverse pages (i.e., those where a single category dominates more than 80\% of the content) dropped by 66\%.
  Likewise, the percentage of users exposed to non-diverse pages decreased by 50\%.
\end{enumerate}

\begin{table}[h]
  \caption{
    A/B test results for integrating query context into the ranking model that powers both the Browse and Search use cases of our e-commerce platform's catalog.
    The table presents the 95\% confidence interval (CI) for the relative uplift of the treatment using Approach B -- query context in the input.
    Engagement is measured based on high-value actions such as add-to-wishlist and add-to-cart.
    Revenue is defined as the net merchandise volume per user after returns.
  }
  \label{tab:zalando-online}
  \centering
  \begin{tabular}{l c c c c}
    \toprule
    & \multicolumn{3}{c}{\textbf{Engagement}} & \\
    \cmidrule(lr){2-4}
    & Browse & Search & All & \textbf{Revenue} \\
    \midrule
    Estimate & +2.4\%       & +0.7\%        & +1.6\%        & +0.7\%  \\
    95\% CI & [2.1, 2.7]\% & [0.3, 1.0]\% & [1.4, 1.9]\% & [0.3, 1.2]\% \\
    \bottomrule
  \end{tabular}
\end{table}

\begin{table*}[t]
  \caption{
    Offline evaluation results when only the current query context is available at inference time.
    The table presents the relative uplift compared to the approach B without query context masking (Section~\ref{sec:fixing-temporal-misalignment}) for three approaches:
    query context outside the transformer (A), query context in the input with the query context masking (B),
    query context in the last layer's query and outside (C). The best performance is denoted in bold font, while the baseline is indicated in underline font.
  }
  \label{tab:open-datasets-no-historical-query-offline}
  \centering
  \begin{tabular}{l c c c c c c c c}
    \toprule
    & \multicolumn{4}{c}{\textbf{Taobao dataset}} & \multicolumn{4}{c}{\textbf{Retailrocket dataset}} \\
    \cmidrule(lr){2-5}
    \cmidrule(lr){6-9}
    \textbf{Query context approach} & Recall@5 & NDCG@5 & Recall@50 & NDCG@50 & Recall@5 & NDCG@5 & Recall@50 & NDCG@50 \\
    \midrule
    Outside the transformer                   & -26.73\% & -33.86\% & -7.97\% & -24.13\% &    -2.2\% & -1.31\% & -0.8\% & -0.69\% \\
    \midrule
    In the input without masking & \underline{0.0929} & \underline{0.0729} & \underline{0.1557} & \underline{0.0888} &     \underline{0.2251} & \underline{0.1629} & \underline{0.6337} & \underline{0.2620} \\
    \midrule
    In the input with masking 0.25 & \textbf{+5.81\%} & \textbf{+6.17\%} & +2.84\% & \textbf{+5.01\%} &     \textbf{+0.62\%} & \textbf{+3.81\%} & -1.43\% & +1.22\% \\
    In the input with masking 0.5  & +1.46\% & +0.29\% & \textbf{+3.09}\% & +1.23\% &     -3.57\% & -1.19\% & -2.98\% & -1.85\% \\
    In the input with masking 0.75 & -1.93\% & -5.16\% & +2.36\% & -2.1\% &      -8.89\% & -5.74\% & -5.45\% & -4.9\% \\
    \midrule
    In the last layer and outside & +5.69\% & +4.59\% & +2.5\% & +3.5\% &       +0.12\% & +2.62\% & \textbf{+0.2\%} & \textbf{+1.51\%} \\
    \bottomrule
  \end{tabular}
\end{table*}

In the second iteration, testing Approach C against the new winner, Approach B,
did not yield statistically significant improvements in the main metrics.
However, it offers advantages in mitigating training-serving mismatches while avoiding unintended misrepresentations,
which we summarize in the following sections.

\section{Experiments using open datasets}

\begin{table*}[t]
  \caption{
    Offline evaluation results showing that the final performance of the environment with historical query context is not superior to the one where only the current query context is available during inference.
    The table presents the relative uplift of the environments using query context in the input (Approach B). The baseline for comparison is indicated in underlined font.
  }
  \label{tab:open-datasets-with-historical-query-offline}
  \centering
  \begin{tabular}{l c c c c c c c c}
    \toprule
    & \multicolumn{4}{c}{\textbf{Taobao dataset}} & \multicolumn{4}{c}{\textbf{Retailrocket dataset}} \\
    \cmidrule(lr){2-5}
    \cmidrule(lr){6-9}
    \textbf{Query context approach} & Recall@5 & NDCG@5 & Recall@50 & NDCG@50 & Recall@5 & NDCG@5 & Recall@50 & NDCG@50 \\
    \midrule
    & \multicolumn{8}{c}{\textbf{Current query context only at inference}} \\
    In the input without masking & \underline{0.0929} & \underline{0.0729} & \underline{0.1557} & \underline{0.0888} &     \underline{0.2251} & \underline{0.1629} & \underline{0.6337} & \underline{0.2620} \\
    In the input with masking 0.25 & +5.81\% & +6.17\% & +2.84\% & +5.01\% &     +0.62\% & +3.81\% & -1.43\% & +1.22\% \\
    \midrule
    & \multicolumn{8}{c}{\textbf{Historical and current query context at inference}} \\
    In the input without masking                      & +5.46\% & +6.18\% & +2.51\% & +4.78\% &       -0.26\% & +1.81\% & +0.25\% & +1.46\% \\
    \bottomrule
  \end{tabular}
\end{table*}

In this section, we evaluate our approaches using two widely used open datasets that include contextual information: \href{https://www.kaggle.com/datasets/marwa80/userbehavior}{Taobao} and \href{https://www.kaggle.com/datasets/retailrocket/ecommerce-dataset}{Retailrocket}.
We follow common preprocessing practices \cite{YANG2021157, liu2024multibehaviorgenerativerecommendation}, filtering out items and categories with fewer than 10 interactions.

\paragraph{Taobao dataset}
This dataset was collected from Taobao, one of the largest e-commerce platforms globally,
and contains user-item interactions from approximately 1 million users over a span of 8 days.
The dataset includes 315,689 items and 7,905 categories.
The median sequence length per user is 46 items, with 0.3\% of sequences consisting of only a single item.

\paragraph{Retailrocket dataset}
This dataset originates from Retail Rocket online platform and consists of 1.5 million users' interactions collected over 4.5 months.
It contains 46,893 unique items and 1,006 unique categories.
A significant portion (70\%) of the user sequences contain only a single interaction, indicating a high prevalence of cold-start cases.

\paragraph{Methodology}
To prevent data leakage, we apply a strict temporal split between the training and test datasets,
where all interactions occurring before a predefined time point are used for training, and those occurring after that time point are reserved for the holdout test set.

We use the next-item category as the query context.
Since the datasets do not specify how often users explicitly select the next-item category (e.g., as a browsing category),
we apply this approximation to all interactions.
We do not include results for models without query context, as they would be inherently disadvantaged in this setting.
However, for reference, on the Taobao dataset, incorporating query context with Approach A led to a 41\% improvement in NDCG@5 compared to a model without query context.

All models are trained until convergence using identical hyperparameters.
We found that using two residual attention-based layers with two heads each was sufficient.

\subsection{Results}

We compare the approaches against the straightforward shifted query context input method (Section~\ref{sec:fixing-temporal-misalignment}).
The results, summarized in Table~\ref{tab:open-datasets-no-historical-query-offline}, show performance when historical query context is available in training,
but only current query context is used in inference.
For Approach B, we conducted a grid search to determine the optimal query context masking rate.
The results show a concave trend, suggesting an optimal masking rate.
We show three masking rate settings, including the best-performing masking rate of 0.25.

As the results suggest, integrating query context into item sequence attention is beneficial.
The best approaches are the optimized Approach B and Approach C, with a marginal difference, where Approach B shows a slight improvement.

Table~\ref{tab:open-datasets-with-historical-query-offline} compares Approach B in two scenarios:
one where only current query context is available during inference and another where both historical and current query context are available.
The results indicate that historical query context in inference does not necessarily improve performance and may not be essential.
Similar experiments conducted on our online platform led to the same conclusion.

The source code, including notebooks for obtaining the datasets, the training procedure, and the model with all variants enabled as feature flags, is available online at \url{https://github.com/djo/query-context-aware-learning-to-rank}.
It also includes details on the hyperparameters used and the implementation.

\section{Discussion}

From our online experiments, integrating query context not only improved engagement and financial metrics but also enhanced recommendation diversity,
spreading visibility and interactions more evenly across the assortment.

\subsection*{{\normalsize RQ1: How can query context be effectively integrated into transformer-based sequential recommendation models while mitigating the training-serving mismatch?}}

Based on experiments with both proprietary and open datasets, the best-performing approaches are Approach B,
which incorporates query context in the input with query context masking,
and Approach C, which applies query context to the last layer's query position as well as outside the transformer.
Both approaches perform comparably, effectively fusing past item representations with query context in attention
while also providing query context as a feature, which is particularly beneficial for cold-start scenarios.

\paragraph{Approach B}
A key advantage of Approach B is that it provides a simple yet effective way to integrate query context
while maintaining compatibility with standard transformer implementations.
It requires minimal modifications, such as adding shifted query context into the input representation and tuning the query context masking via grid search.
This makes it a practical choice, especially when using off-the-shelf transformer models.

However, there are some limitations, as highlighted in Section~\ref{sec:fixing-temporal-misalignment}:
\begin{enumerate}
  \item \textbf{Training-serving mismatch:}
  When historical query context is unavailable during inference, a discrepancy remains between training and serving, despite the introduction of query context masking.
  \item \textbf{Challenges in learning and representation:}
  Since each past item is represented alongside the next-item query context rather than its intrinsic features,
  learning may become unnecessarily complex or less generalizable, especially in sparse data settings.
  This issue requires further investigation, particularly when the query context becomes more feature-rich and includes diverse signals.
  \item \textbf{Limited extensibility:}
  The approach may be less suited for integration with interleaving features, such as methods using random time window masking \cite{10.1145/3580305.3599918}.
\end{enumerate}

\paragraph{Approach C}
A key advantage of Approach C is that it remains robust against training-serving mismatches and avoids issues related to misleading representations and extensibility.
A potential drawback is that this approach does not leverage historical query context, as we hypothesize that in some datasets, it could be beneficial when historical query context is available online.
Having a sequence of query contexts as an additional signal might improve performance.
However, this limitation could be mitigated by using item features as contextual information in the input.
For example, if the query context is a browsing category, item categories could serve as contextual features,
providing a sequence of categories as the necessary additional signal.
This remains an open question for future research.

\subsection*{{\normalsize RQ2: Does the absence of historical query context during inference negatively impact model performance?}}

To determine whether having historical query context in inference is beneficial, we conduct an offline evaluation using proprietary and open-source datasets where both historical and current query context are available during inference,
using Approach B, as it is the only approach that might benefit from this information.
Experiments on these datasets show that incorporating historical query context does not necessarily improve performance, suggesting that its availability online may not be essential.

\section{Conclusion}

This paper introduces effective and efficient approaches for integrating query context into transformer-based recommendation models.
We validate their effectiveness through extensive offline and online experiments on a large-scale online platform and open datasets.
Our experiments demonstrate that the proposed methods improve ranking quality in terms of both relevance and diversity.
The most efficient approaches either incorporate query context into the input or
apply it to the query position in the last layer, as well as outside the transformer.
The latter approach completely avoids the training-serving mismatch when historical query context is unavailable.
Our findings suggest that having historical query context available online may not be essential for achieving strong performance.

\section*{Acknowledgement}

We are grateful for the valuable feedback, insightful discussions, and constant support from our many colleagues, as well as their contributions to the design and execution of the online experiments, including
Geraud Le Falhe, Danilo Ascione, Tural Gurbanov, Egor Malykh, Gabriel Coelho, Shashi Kumar, Tao Ruangyam, and John Coleman.

\bibliographystyle{ACM-Reference-Format}
\bibliography{main}


\begin{thebibliography}{29}


\ifx \showCODEN    \undefined \def \showCODEN     #1{\unskip}     \fi
\ifx \showISBNx    \undefined \def \showISBNx     #1{\unskip}     \fi
\ifx \showISBNxiii \undefined \def \showISBNxiii  #1{\unskip}     \fi
\ifx \showISSN     \undefined \def \showISSN      #1{\unskip}     \fi
\ifx \showLCCN     \undefined \def \showLCCN      #1{\unskip}     \fi
\ifx \shownote     \undefined \def \shownote      #1{#1}          \fi
\ifx \showarticletitle \undefined \def \showarticletitle #1{#1}   \fi
\ifx \showURL      \undefined \def \showURL       {\relax}        \fi
\providecommand\bibfield[2]{#2}
\providecommand\bibinfo[2]{#2}
\providecommand\natexlab[1]{#1}
\providecommand\showeprint[2][]{arXiv:#2}

\bibitem[Adomavicius and Tuzhilin(2015)]%
        {Adomavicius2015}
\bibfield{author}{\bibinfo{person}{Gediminas Adomavicius} {and}
  \bibinfo{person}{Alexander Tuzhilin}.} \bibinfo{year}{2015}\natexlab{}.
\newblock \bibinfo{booktitle}{\emph{Context-Aware Recommender Systems}}.
\newblock \bibinfo{publisher}{Springer US}, \bibinfo{address}{Boston, MA},
  \bibinfo{pages}{191--226}.
\newblock
\showISBNx{978-1-4899-7637-6}
\href{https://doi.org/10.1007/978-1-4899-7637-6_6}{doi:\nolinkurl{10.1007/978-1-4899-7637-6_6}}


\bibitem[Ba et~al\mbox{.}(2016)]%
        {ba2016layer}
\bibfield{author}{\bibinfo{person}{Jimmy~Lei Ba}, \bibinfo{person}{Jamie~Ryan
  Kiros}, {and} \bibinfo{person}{Geoffrey~E Hinton}.}
  \bibinfo{year}{2016}\natexlab{}.
\newblock \showarticletitle{Layer normalization}.
\newblock \bibinfo{journal}{\emph{arXiv preprint arXiv:1607.06450}}
  (\bibinfo{year}{2016}).
\newblock


\bibitem[Beutel et~al\mbox{.}(2018)]%
        {10.1145/3159652.3159727}
\bibfield{author}{\bibinfo{person}{Alex Beutel}, \bibinfo{person}{Paul
  Covington}, \bibinfo{person}{Sagar Jain}, \bibinfo{person}{Can Xu},
  \bibinfo{person}{Jia Li}, \bibinfo{person}{Vince Gatto}, {and}
  \bibinfo{person}{Ed~H. Chi}.} \bibinfo{year}{2018}\natexlab{}.
\newblock \showarticletitle{Latent Cross: Making Use of Context in Recurrent
  Recommender Systems}. In \bibinfo{booktitle}{\emph{Proceedings of the
  Eleventh ACM International Conference on Web Search and Data Mining}} (Marina
  Del Rey, CA, USA) \emph{(\bibinfo{series}{WSDM '18})}.
  \bibinfo{publisher}{Association for Computing Machinery},
  \bibinfo{address}{New York, NY, USA}, \bibinfo{pages}{46–54}.
\newblock
\showISBNx{9781450355810}
\href{https://doi.org/10.1145/3159652.3159727}{doi:\nolinkurl{10.1145/3159652.3159727}}


\bibitem[Cai et~al\mbox{.}(2021)]%
        {10.1145/3404835.3462832}
\bibfield{author}{\bibinfo{person}{Renqin Cai}, \bibinfo{person}{Jibang Wu},
  \bibinfo{person}{Aidan San}, \bibinfo{person}{Chong Wang}, {and}
  \bibinfo{person}{Hongning Wang}.} \bibinfo{year}{2021}\natexlab{}.
\newblock \showarticletitle{Category-aware Collaborative Sequential
  Recommendation}. In \bibinfo{booktitle}{\emph{Proceedings of the 44th
  International ACM SIGIR Conference on Research and Development in Information
  Retrieval}} (Virtual Event, Canada) \emph{(\bibinfo{series}{SIGIR '21})}.
  \bibinfo{publisher}{Association for Computing Machinery},
  \bibinfo{address}{New York, NY, USA}, \bibinfo{pages}{388–397}.
\newblock
\showISBNx{9781450380379}
\href{https://doi.org/10.1145/3404835.3462832}{doi:\nolinkurl{10.1145/3404835.3462832}}


\bibitem[Cao et~al\mbox{.}(2008)]%
        {10.1145/1401890.1401995}
\bibfield{author}{\bibinfo{person}{Huanhuan Cao}, \bibinfo{person}{Daxin
  Jiang}, \bibinfo{person}{Jian Pei}, \bibinfo{person}{Qi He},
  \bibinfo{person}{Zhen Liao}, \bibinfo{person}{Enhong Chen}, {and}
  \bibinfo{person}{Hang Li}.} \bibinfo{year}{2008}\natexlab{}.
\newblock \showarticletitle{Context-aware query suggestion by mining
  click-through and session data}. In \bibinfo{booktitle}{\emph{Proceedings of
  the 14th ACM SIGKDD International Conference on Knowledge Discovery and Data
  Mining}} (Las Vegas, Nevada, USA) \emph{(\bibinfo{series}{KDD '08})}.
  \bibinfo{publisher}{Association for Computing Machinery},
  \bibinfo{address}{New York, NY, USA}, \bibinfo{pages}{875–883}.
\newblock
\showISBNx{9781605581934}
\href{https://doi.org/10.1145/1401890.1401995}{doi:\nolinkurl{10.1145/1401890.1401995}}


\bibitem[Celikik et~al\mbox{.}(2024)]%
        {celikik2024buildingscalableeffectivesteerable}
\bibfield{author}{\bibinfo{person}{Marjan Celikik}, \bibinfo{person}{Jacek
  Wasilewski}, \bibinfo{person}{Ana~Peleteiro Ramallo}, \bibinfo{person}{Alexey
  Kurennoy}, \bibinfo{person}{Evgeny Labzin}, \bibinfo{person}{Danilo Ascione},
  \bibinfo{person}{Tural Gurbanov}, \bibinfo{person}{Géraud~Le Falher},
  \bibinfo{person}{Andrii Dzhoha}, {and} \bibinfo{person}{Ian Harris}.}
  \bibinfo{year}{2024}\natexlab{}.
\newblock \bibinfo{title}{Building a Scalable, Effective, and Steerable Search
  and Ranking Platform}.
\newblock
\showeprint[arxiv]{2409.02856}~[cs.IR]
\urldef\tempurl%
\url{https://arxiv.org/abs/2409.02856}
\showURL{%
\tempurl}


\bibitem[Covington et~al\mbox{.}(2016)]%
        {10.1145/2959100.2959190}
\bibfield{author}{\bibinfo{person}{Paul Covington}, \bibinfo{person}{Jay
  Adams}, {and} \bibinfo{person}{Emre Sargin}.}
  \bibinfo{year}{2016}\natexlab{}.
\newblock \showarticletitle{Deep Neural Networks for YouTube Recommendations}.
  In \bibinfo{booktitle}{\emph{Proceedings of the 10th ACM Conference on
  Recommender Systems}} (Boston, Massachusetts, USA)
  \emph{(\bibinfo{series}{RecSys '16})}. \bibinfo{publisher}{Association for
  Computing Machinery}, \bibinfo{address}{New York, NY, USA},
  \bibinfo{pages}{191–198}.
\newblock
\showISBNx{9781450340359}
\href{https://doi.org/10.1145/2959100.2959190}{doi:\nolinkurl{10.1145/2959100.2959190}}


\bibitem[Di~Teodoro et~al\mbox{.}(2024)]%
        {di2024theoretical}
\bibfield{author}{\bibinfo{person}{Giulia Di~Teodoro},
  \bibinfo{person}{Federico Siciliano}, \bibinfo{person}{Nicola Tonellotto},
  {and} \bibinfo{person}{Fabrizio Silvestri}.} \bibinfo{year}{2024}\natexlab{}.
\newblock \showarticletitle{A Theoretical Analysis of Recommendation Loss
  Functions under Negative Sampling}.
\newblock \bibinfo{journal}{\emph{arXiv preprint arXiv:2411.07770}}
  (\bibinfo{year}{2024}).
\newblock


\bibitem[Dzhoha et~al\mbox{.}(2024)]%
        {dzhoha2024reducingpopularityinfluenceaddressing}
\bibfield{author}{\bibinfo{person}{Andrii Dzhoha}, \bibinfo{person}{Alexey
  Kurennoy}, \bibinfo{person}{Vladimir Vlasov}, {and} \bibinfo{person}{Marjan
  Celikik}.} \bibinfo{year}{2024}\natexlab{}.
\newblock \bibinfo{title}{Reducing Popularity Influence by Addressing Position
  Bias}.
\newblock
\showeprint[arxiv]{2412.08780}~[cs.IR]
\urldef\tempurl%
\url{https://arxiv.org/abs/2412.08780}
\showURL{%
\tempurl}


\bibitem[Kang and McAuley(2018)]%
        {Kang2018SelfAttentiveSR}
\bibfield{author}{\bibinfo{person}{Wang-Cheng Kang} {and}
  \bibinfo{person}{Julian McAuley}.} \bibinfo{year}{2018}\natexlab{}.
\newblock \showarticletitle{Self-Attentive Sequential Recommendation}.
\newblock \bibinfo{journal}{\emph{2018 IEEE International Conference on Data
  Mining (ICDM)}} (\bibinfo{year}{2018}), \bibinfo{pages}{197--206}.
\newblock


\bibitem[Kang et~al\mbox{.}(2024)]%
        {10711191}
\bibfield{author}{\bibinfo{person}{Woo-Seung Kang}, \bibinfo{person}{Hye-Jin
  Jeong}, \bibinfo{person}{Suwon Lee}, {and} \bibinfo{person}{Sang-Min Choi}.}
  \bibinfo{year}{2024}\natexlab{}.
\newblock \showarticletitle{Independent Representation of Side Information for
  Sequential Recommendation}.
\newblock \bibinfo{journal}{\emph{IEEE Access}}  \bibinfo{volume}{12}
  (\bibinfo{year}{2024}), \bibinfo{pages}{148516--148524}.
\newblock
\href{https://doi.org/10.1109/ACCESS.2024.3476976}{doi:\nolinkurl{10.1109/ACCESS.2024.3476976}}


\bibitem[Li et~al\mbox{.}(2020)]%
        {10.1145/3336191.3371786}
\bibfield{author}{\bibinfo{person}{Jiacheng Li}, \bibinfo{person}{Yujie Wang},
  {and} \bibinfo{person}{Julian McAuley}.} \bibinfo{year}{2020}\natexlab{}.
\newblock \showarticletitle{Time Interval Aware Self-Attention for Sequential
  Recommendation}. In \bibinfo{booktitle}{\emph{Proceedings of the 13th
  International Conference on Web Search and Data Mining}} (Houston, TX, USA)
  \emph{(\bibinfo{series}{WSDM '20})}. \bibinfo{publisher}{Association for
  Computing Machinery}, \bibinfo{address}{New York, NY, USA},
  \bibinfo{pages}{322–330}.
\newblock
\showISBNx{9781450368223}
\href{https://doi.org/10.1145/3336191.3371786}{doi:\nolinkurl{10.1145/3336191.3371786}}


\bibitem[Li et~al\mbox{.}(2017)]%
        {pmlr-v67-li17a}
\bibfield{author}{\bibinfo{person}{Li~Erran Li}, \bibinfo{person}{Eric Chen},
  \bibinfo{person}{Jeremy Hermann}, \bibinfo{person}{Pusheng Zhang}, {and}
  \bibinfo{person}{Luming Wang}.} \bibinfo{year}{2017}\natexlab{}.
\newblock \showarticletitle{Scaling Machine Learning as a Service}. In
  \bibinfo{booktitle}{\emph{Proceedings of The 3rd International Conference on
  Predictive Applications and APIs}} \emph{(\bibinfo{series}{Proceedings of
  Machine Learning Research}, Vol.~\bibinfo{volume}{67})},
  \bibfield{editor}{\bibinfo{person}{Claire Hardgrove}, \bibinfo{person}{Louis
  Dorard}, \bibinfo{person}{Keiran Thompson}, {and} \bibinfo{person}{Florian
  Douetteau}} (Eds.). \bibinfo{publisher}{PMLR}, \bibinfo{pages}{14--29}.
\newblock
\urldef\tempurl%
\url{https://proceedings.mlr.press/v67/li17a.html}
\showURL{%
\tempurl}


\bibitem[Liu et~al\mbox{.}(2016)]%
        {7837948}
\bibfield{author}{\bibinfo{person}{Qiang Liu}, \bibinfo{person}{Shu Wu},
  \bibinfo{person}{Diyi Wang}, \bibinfo{person}{Zhaokang Li}, {and}
  \bibinfo{person}{Liang Wang}.} \bibinfo{year}{2016}\natexlab{}.
\newblock \showarticletitle{Context-Aware Sequential Recommendation}. In
  \bibinfo{booktitle}{\emph{2016 IEEE 16th International Conference on Data
  Mining (ICDM)}}. \bibinfo{pages}{1053--1058}.
\newblock
\href{https://doi.org/10.1109/ICDM.2016.0135}{doi:\nolinkurl{10.1109/ICDM.2016.0135}}


\bibitem[Liu et~al\mbox{.}(2024)]%
        {liu2024multibehaviorgenerativerecommendation}
\bibfield{author}{\bibinfo{person}{Zihan Liu}, \bibinfo{person}{Yupeng Hou},
  {and} \bibinfo{person}{Julian McAuley}.} \bibinfo{year}{2024}\natexlab{}.
\newblock \showarticletitle{Multi-Behavior Generative Recommendation}.
\newblock  (\bibinfo{year}{2024}).
\newblock
\showeprint[arxiv]{2405.16871}~[cs.IR]
\urldef\tempurl%
\url{https://arxiv.org/abs/2405.16871}
\showURL{%
\tempurl}


\bibitem[Nguyen and Salazar(2019)]%
        {nguyen2019transformers}
\bibfield{author}{\bibinfo{person}{Toan~Q Nguyen} {and} \bibinfo{person}{Julian
  Salazar}.} \bibinfo{year}{2019}\natexlab{}.
\newblock \showarticletitle{Transformers without tears: Improving the
  normalization of self-attention}.
\newblock \bibinfo{journal}{\emph{arXiv preprint arXiv:1910.05895}}
  (\bibinfo{year}{2019}).
\newblock


\bibitem[Quadrana et~al\mbox{.}(2018)]%
        {10.1145/3190616}
\bibfield{author}{\bibinfo{person}{Massimo Quadrana}, \bibinfo{person}{Paolo
  Cremonesi}, {and} \bibinfo{person}{Dietmar Jannach}.}
  \bibinfo{year}{2018}\natexlab{}.
\newblock \showarticletitle{Sequence-Aware Recommender Systems}.
\newblock \bibinfo{journal}{\emph{ACM Comput. Surv.}} \bibinfo{volume}{51},
  \bibinfo{number}{4}, Article \bibinfo{articleno}{66} (\bibinfo{date}{July}
  \bibinfo{year}{2018}), \bibinfo{numpages}{36}~pages.
\newblock
\showISSN{0360-0300}
\href{https://doi.org/10.1145/3190616}{doi:\nolinkurl{10.1145/3190616}}


\bibitem[Song et~al\mbox{.}(2017)]%
        {7917325}
\bibfield{author}{\bibinfo{person}{Jun Song}, \bibinfo{person}{Jun Xiao},
  \bibinfo{person}{Fei Wu}, \bibinfo{person}{Haishan Wu}, \bibinfo{person}{Tong
  Zhang}, \bibinfo{person}{Zhongfei~Mark Zhang}, {and} \bibinfo{person}{Wenwu
  Zhu}.} \bibinfo{year}{2017}\natexlab{}.
\newblock \showarticletitle{Hierarchical Contextual Attention Recurrent Neural
  Network for Map Query Suggestion}.
\newblock \bibinfo{journal}{\emph{IEEE Transactions on Knowledge and Data
  Engineering}} \bibinfo{volume}{29}, \bibinfo{number}{9}
  (\bibinfo{year}{2017}), \bibinfo{pages}{1888--1901}.
\newblock
\href{https://doi.org/10.1109/TKDE.2017.2700392}{doi:\nolinkurl{10.1109/TKDE.2017.2700392}}


\bibitem[Sun et~al\mbox{.}(2019)]%
        {10.1145/3357384.3357895}
\bibfield{author}{\bibinfo{person}{Fei Sun}, \bibinfo{person}{Jun Liu},
  \bibinfo{person}{Jian Wu}, \bibinfo{person}{Changhua Pei},
  \bibinfo{person}{Xiao Lin}, \bibinfo{person}{Wenwu Ou}, {and}
  \bibinfo{person}{Peng Jiang}.} \bibinfo{year}{2019}\natexlab{}.
\newblock \showarticletitle{BERT4Rec: Sequential Recommendation with
  Bidirectional Encoder Representations from Transformer}. In
  \bibinfo{booktitle}{\emph{Proceedings of the 28th ACM International
  Conference on Information and Knowledge Management}} (Beijing, China)
  \emph{(\bibinfo{series}{CIKM '19})}. \bibinfo{publisher}{Association for
  Computing Machinery}, \bibinfo{address}{New York, NY, USA},
  \bibinfo{pages}{1441–1450}.
\newblock
\showISBNx{9781450369763}
\href{https://doi.org/10.1145/3357384.3357895}{doi:\nolinkurl{10.1145/3357384.3357895}}


\bibitem[Tamm et~al\mbox{.}(2021)]%
        {10.1145/3460231.3478848}
\bibfield{author}{\bibinfo{person}{Yan-Martin Tamm}, \bibinfo{person}{Rinchin
  Damdinov}, {and} \bibinfo{person}{Alexey Vasilev}.}
  \bibinfo{year}{2021}\natexlab{}.
\newblock \showarticletitle{Quality Metrics in Recommender Systems: Do We
  Calculate Metrics Consistently?}. In \bibinfo{booktitle}{\emph{Proceedings of
  the 15th ACM Conference on Recommender Systems}} (Amsterdam, Netherlands)
  \emph{(\bibinfo{series}{RecSys '21})}. \bibinfo{publisher}{Association for
  Computing Machinery}, \bibinfo{address}{New York, NY, USA},
  \bibinfo{pages}{708–713}.
\newblock
\showISBNx{9781450384582}
\href{https://doi.org/10.1145/3460231.3478848}{doi:\nolinkurl{10.1145/3460231.3478848}}


\bibitem[Vaswani et~al\mbox{.}(2017)]%
        {10.5555/3295222.3295349}
\bibfield{author}{\bibinfo{person}{Ashish Vaswani}, \bibinfo{person}{Noam
  Shazeer}, \bibinfo{person}{Niki Parmar}, \bibinfo{person}{Jakob Uszkoreit},
  \bibinfo{person}{Llion Jones}, \bibinfo{person}{Aidan~N. Gomez},
  \bibinfo{person}{\L{}ukasz Kaiser}, {and} \bibinfo{person}{Illia
  Polosukhin}.} \bibinfo{year}{2017}\natexlab{}.
\newblock \showarticletitle{Attention is all you need}. In
  \bibinfo{booktitle}{\emph{Proceedings of the 31st International Conference on
  Neural Information Processing Systems}} (Long Beach, California, USA)
  \emph{(\bibinfo{series}{NIPS'17})}. \bibinfo{publisher}{Curran Associates
  Inc.}, \bibinfo{address}{Red Hook, NY, USA}, \bibinfo{pages}{6000–6010}.
\newblock
\showISBNx{9781510860964}


\bibitem[Vlasov(2024)]%
        {2401.09343}
\bibfield{author}{\bibinfo{person}{Vladimir Vlasov}.}
  \bibinfo{year}{2024}\natexlab{}.
\newblock \bibinfo{title}{Efficient slot labelling}.
\newblock
\showeprint[arxiv]{2401.09343}~[cs.CL]
\urldef\tempurl%
\url{https://arxiv.org/abs/2401.09343}
\showURL{%
\tempurl}


\bibitem[Xia et~al\mbox{.}(2023)]%
        {10.1145/3580305.3599918}
\bibfield{author}{\bibinfo{person}{Xue Xia}, \bibinfo{person}{Pong
  Eksombatchai}, \bibinfo{person}{Nikil Pancha}, \bibinfo{person}{Dhruvil~Deven
  Badani}, \bibinfo{person}{Po-Wei Wang}, \bibinfo{person}{Neng Gu},
  \bibinfo{person}{Saurabh~Vishwas Joshi}, \bibinfo{person}{Nazanin Farahpour},
  \bibinfo{person}{Zhiyuan Zhang}, {and} \bibinfo{person}{Andrew Zhai}.}
  \bibinfo{year}{2023}\natexlab{}.
\newblock \showarticletitle{TransAct: Transformer-based Realtime User Action
  Model for Recommendation at Pinterest}. In
  \bibinfo{booktitle}{\emph{Proceedings of the 29th ACM SIGKDD Conference on
  Knowledge Discovery and Data Mining}} (Long Beach, CA, USA)
  \emph{(\bibinfo{series}{KDD '23})}. \bibinfo{publisher}{Association for
  Computing Machinery}, \bibinfo{address}{New York, NY, USA},
  \bibinfo{pages}{5249–5259}.
\newblock
\showISBNx{9798400701030}
\href{https://doi.org/10.1145/3580305.3599918}{doi:\nolinkurl{10.1145/3580305.3599918}}


\bibitem[Xie et~al\mbox{.}(2022)]%
        {10.1145/3477495.3531963}
\bibfield{author}{\bibinfo{person}{Yueqi Xie}, \bibinfo{person}{Peilin Zhou},
  {and} \bibinfo{person}{Sunghun Kim}.} \bibinfo{year}{2022}\natexlab{}.
\newblock \showarticletitle{Decoupled Side Information Fusion for Sequential
  Recommendation}. In \bibinfo{booktitle}{\emph{Proceedings of the 45th
  International ACM SIGIR Conference on Research and Development in Information
  Retrieval}} (Madrid, Spain) \emph{(\bibinfo{series}{SIGIR '22})}.
  \bibinfo{publisher}{Association for Computing Machinery},
  \bibinfo{address}{New York, NY, USA}, \bibinfo{pages}{1611–1621}.
\newblock
\showISBNx{9781450387323}
\href{https://doi.org/10.1145/3477495.3531963}{doi:\nolinkurl{10.1145/3477495.3531963}}


\bibitem[Yang et~al\mbox{.}(2021)]%
        {YANG2021157}
\bibfield{author}{\bibinfo{person}{Baosong Yang}, \bibinfo{person}{Longyue
  Wang}, \bibinfo{person}{Derek~F. Wong}, \bibinfo{person}{Shuming Shi}, {and}
  \bibinfo{person}{Zhaopeng Tu}.} \bibinfo{year}{2021}\natexlab{}.
\newblock \showarticletitle{Context-aware Self-Attention Networks for Natural
  Language Processing}.
\newblock \bibinfo{journal}{\emph{Neurocomputing}}  \bibinfo{volume}{458}
  (\bibinfo{year}{2021}), \bibinfo{pages}{157--169}.
\newblock
\showISSN{0925-2312}
\href{https://doi.org/10.1016/j.neucom.2021.06.009}{doi:\nolinkurl{10.1016/j.neucom.2021.06.009}}


\bibitem[Yang et~al\mbox{.}(2020)]%
        {twotowersampling}
\bibfield{author}{\bibinfo{person}{Ji Yang}, \bibinfo{person}{Xinyang Yi},
  \bibinfo{person}{Derek Zhiyuan~Cheng}, \bibinfo{person}{Lichan Hong},
  \bibinfo{person}{Yang Li}, \bibinfo{person}{Simon Xiaoming~Wang},
  \bibinfo{person}{Taibai Xu}, {and} \bibinfo{person}{Ed~H. Chi}.}
  \bibinfo{year}{2020}\natexlab{}.
\newblock \showarticletitle{Mixed Negative Sampling for Learning Two-tower
  Neural Networks in Recommendations}. In \bibinfo{booktitle}{\emph{Companion
  Proceedings of the Web Conference 2020}} (Taipei, Taiwan)
  \emph{(\bibinfo{series}{WWW '20})}. \bibinfo{publisher}{Association for
  Computing Machinery}, \bibinfo{address}{New York, NY, USA},
  \bibinfo{pages}{441–447}.
\newblock
\showISBNx{9781450370240}
\href{https://doi.org/10.1145/3366424.3386195}{doi:\nolinkurl{10.1145/3366424.3386195}}


\bibitem[Zhang et~al\mbox{.}(2017)]%
        {8031316}
\bibfield{author}{\bibinfo{person}{Biao Zhang}, \bibinfo{person}{Deyi Xiong},
  \bibinfo{person}{Jinsong Su}, {and} \bibinfo{person}{Hong Duan}.}
  \bibinfo{year}{2017}\natexlab{}.
\newblock \showarticletitle{A Context-Aware Recurrent Encoder for Neural
  Machine Translation}.
\newblock \bibinfo{journal}{\emph{IEEE/ACM Transactions on Audio, Speech, and
  Language Processing}} \bibinfo{volume}{25}, \bibinfo{number}{12}
  (\bibinfo{year}{2017}), \bibinfo{pages}{2424--2432}.
\newblock
\href{https://doi.org/10.1109/TASLP.2017.2751420}{doi:\nolinkurl{10.1109/TASLP.2017.2751420}}


\bibitem[Zhang et~al\mbox{.}(2019)]%
        {10.5555/3367471.3367642}
\bibfield{author}{\bibinfo{person}{Tingting Zhang}, \bibinfo{person}{Pengpeng
  Zhao}, \bibinfo{person}{Yanchi Liu}, \bibinfo{person}{Victor~S. Sheng},
  \bibinfo{person}{Jiajie Xu}, \bibinfo{person}{Deqing Wang},
  \bibinfo{person}{Guanfeng Liu}, {and} \bibinfo{person}{Xiaofang Zhou}.}
  \bibinfo{year}{2019}\natexlab{}.
\newblock \showarticletitle{Feature-level deeper self-attention network for
  sequential recommendation}. In \bibinfo{booktitle}{\emph{Proceedings of the
  28th International Joint Conference on Artificial Intelligence}} (Macao,
  China) \emph{(\bibinfo{series}{IJCAI'19})}. \bibinfo{publisher}{AAAI Press},
  \bibinfo{pages}{4320–4326}.
\newblock
\showISBNx{9780999241141}


\bibitem[Zhou et~al\mbox{.}(2018)]%
        {10.1145/3219819.3219823}
\bibfield{author}{\bibinfo{person}{Guorui Zhou}, \bibinfo{person}{Xiaoqiang
  Zhu}, \bibinfo{person}{Chenru Song}, \bibinfo{person}{Ying Fan},
  \bibinfo{person}{Han Zhu}, \bibinfo{person}{Xiao Ma},
  \bibinfo{person}{Yanghui Yan}, \bibinfo{person}{Junqi Jin},
  \bibinfo{person}{Han Li}, {and} \bibinfo{person}{Kun Gai}.}
  \bibinfo{year}{2018}\natexlab{}.
\newblock \showarticletitle{Deep Interest Network for Click-Through Rate
  Prediction}. In \bibinfo{booktitle}{\emph{Proceedings of the 24th ACM SIGKDD
  International Conference on Knowledge Discovery \& Data Mining}} (London,
  United Kingdom) \emph{(\bibinfo{series}{KDD '18})}.
  \bibinfo{publisher}{Association for Computing Machinery},
  \bibinfo{address}{New York, NY, USA}, \bibinfo{pages}{1059–1068}.
\newblock
\showISBNx{9781450355520}
\href{https://doi.org/10.1145/3219819.3219823}{doi:\nolinkurl{10.1145/3219819.3219823}}


\end{thebibliography}

\end{document}